\documentclass[letterpaper, 10 pt, conference]{ieeeconf}  

\IEEEoverridecommandlockouts                              
\usepackage{graphics} 
\usepackage{epsfig} 
\usepackage{times} 
\usepackage{amsmath} 
\usepackage{amssymb}  

\title{\LARGE \bf
Quantum state transfer for multi-input linear quantum systems
}

\author{Naoki Yamamoto, Hendra I. Nurdin, and Matthew R. James
\thanks{N. Yamamoto is with the Department of Applied Physics and 
               Physico-Informatics, Keio University, Hiyoshi 3-14-1, Kohoku, 
               Yokohama, Japan (e-mail: yamamoto@appi.keio.ac.jp). 
               H. I. Nurdin is with School of Electrical Engineering and 
               Telecommunications, UNSW Australia, Sydney NSW 2052, 
               Australia (e-mail: h.nurdin@unsw.edu.au). 
               M. R. James is with Research School of Engineering, The Australian 
               National University, Canberra, ACT 0200, Australia 
               (e-mail: Matthew.James@anu.edu.au). 
               This work is supported by the Australian Research Council, the 
               UNSW Australia Faculty of Engineering, and JSPS Grant-in-Aid 
               No. 15K06151. 
}
}

\usepackage{eclbkbox}

\newcommand{\pro}[2]{\langle{#1}|{#2}\rangle}
\newcommand{\bra}[1]{\langle{#1}|}
\newcommand{\ket}[1]{|{#1}\rangle}

\newcommand{\veca}{\mbox{\boldmath $a$}}
\newcommand{\vecb}{\mbox{\boldmath $b$}}
\newcommand{\vecy}{\mbox{\boldmath $y$}}
\newcommand{\vecB}{\mbox{\boldmath $B$}}
\newcommand{\vecu}{\mbox{\boldmath $u$}}
\newcommand{\vecv}{\mbox{\boldmath $v$}}

\begin{document}

\maketitle
\thispagestyle{empty}
\pagestyle{empty}


\begin{abstract}
Effective state transfer is one of the most important problems in quantum 
information processing. 
Typically, a quantum information device is composed of many subsystems 
with multi-input ports. 
In this paper, we develop a general theory describing the condition for perfect 
state transfer from the multi-input ports to the internal system components, 
for general passive linear quantum systems. 
The key notion used is the zero of the transfer function matrix. 
Application to entanglement generation and distribution in a quantum network 
is also discussed. 
\end{abstract}


\section{Introduction}

A quantum state transducer, that for instance transfers an optical state to 
a solid state, is an indispensable component contained in various types of 
quantum information processors. 
For instance, such a state transfer procedure is involved in every quantum 
memory architecture \cite{PrebleOptExpress2010,Painter 2011 memory,
Polzik2011,Focus on}, which is typically used for building a quantum 
repeater in quantum communication networks \cite{Repeater review}. 
Towards a systematic and effective design of state transfer protocol, in 
\cite{Yamamoto 2014} two of the authors developed a general theory for 
single-input and single-output (SISO) passive linear quantum systems 
\cite{GoughPRA2008,Guta Yamamoto}; 
the result obtained is that the input state encoded in an appropriately shaped 
wave function can be perfectly transferred to the system. 
A remarkable fact is that such a wave function can be completely characterized 
in terms of the zeros of the transfer function of the system, which thus revealed 
a close connection between systems and control theory and the important 
task in quantum information science.

Based on this background, in this paper, we aim to extend the result of 
\cite{Yamamoto 2014} to the case of multi-input and multi-output (MIMO) 
general linear passive systems. 
In fact the memory systems studied in the above-referred papers 
\cite{PrebleOptExpress2010,Painter 2011 memory,Polzik2011} are all 
MIMO systems. 
Also a hybridized system conducting frequency conversion between e.g. 
an optical cavity and a microwave circuit is essentially an MIMO system 
\cite{Painter 2011,Clerk 2012,Nakamura 2016}. 
On the other hand, it is well known in classical systems and control theory 
\cite{Zhou Doyle book} that extending the notion of zeros from the SISO case 
to the MIMO case is quite nontrivial. 
This is because in the MIMO case we are dealing with a transfer function 
{\it matrix}, and the zeros of this matrix can have several definitions; 
for instance, a {\it transmission zero} is defined as a complex number at which 
the rank of the transfer function matrix drops, while a {\it blocking zero} 
is a complex number at which the transfer function matrix becomes a zero 
matrix. 
Therefore, the goal of this paper is to deduce the condition for perfect state 
transfer and how that condition can be characterized by the zeros of the 
transfer function matrix.

{\it Notation:} 
for a matrix $A=(a_{ij})$, the symbols $A^\dagger$, $A^\top$, and 
$A^\sharp$ represent its Hermitian conjugate, transpose, and complex 
conjugation in elements of $A$, respectively; i.e., 
$A^\dagger=(a_{ji}^*)$, $A^\top=(a_{ji})$, and $A^\sharp=(a^*_{ij})$. 
For a matrix of operators we use the same notation, in which case $a_{ij}^*$ 
denotes the adjoint to $a_{ij}$.


\section{Preliminaries}


\subsection{Model of the system and input}

Let us consider the following MIMO passive linear quantum system 
\cite{GoughPRA2008,Guta Yamamoto}: 
\begin{equation}
\label{memory dynamics}
      \frac{d\veca}{dt}=A\veca - C^\dagger S\vecb,~~~
      \tilde{\vecb}=C\veca + S\vecb.
\end{equation}
Here $\veca=[a_1, \ldots, a_n]^\top$ is the vector of system annihilation 
operators. 
This system has $m$ input channels represented by the vector of field 
annihilation operators $\vecb=[b_1, \ldots, b_m]^\top$, and $\tilde{\vecb}$ 
is the corresponding output. 
These are infinite dimensional operators satisfying e.g. 
$a_i(t)a_j^*(t)-a_j^*(t)a_i(t)=\delta_{ij}~\forall t$ and 
$b_i(t)b_j^*(t')-b_j^*(t')b_i(t)=\delta_{ij}\delta(t-t')~\forall t, t'$. 
In the dynamical equation, $C\in{\mathbb C}^{m\times n}$ represents the 
system-field coupling; 
also $A=-i\Omega -C^\dagger C/2$, where the $n\times n$ Hermitian matrix 
$\Omega=\Omega^\dagger$ is related to the system Hamiltonian. 
Finally $S$ is a $m\times m$ unitary matrix, representing the scattering 
process of $\vecb$.

In the state transfer problem considered in this paper, we assume that 
the input is given by a continuous-mode single-photon field state. 
This state is defined in terms of the following annihilation and creation 
process operators: 
\begin{equation}
\label{B and Bstar}
       B(\xi) = \int_{-\infty}^\infty \xi^*(t)b(t)dt,~~
       B^*(\xi) = \int_{-\infty}^\infty \xi(t)b^* (t)dt. 
\end{equation}
$\xi(t)$ is an associated function in ${\mathbb C}$, representing the shape 
of the optical pulse field. 
Also $\xi(t)$ satisfies the normalization condition 
$\int_{-\infty}^\infty |\xi(t)|^2 dt=1$. 
Due to this, $B(\xi)$ and $B^*(\xi)$ satisfy the relation 
$B(\xi)B^*(\xi) - B^*(\xi)B(\xi)=1$. 
The continuous-mode single photon field state is produced by acting $B^*(\xi)$ 
on the vacuum field $\ket{0}_f$ as follows \cite{Zoller 1998,Milburn 2008}: 
\begin{equation}
\label{cont mode single photon}
   \ket{1_\xi}_f = B^*(\xi)\ket{0}_f 
           = \int_{-\infty}^\infty \xi(t)b^* (t)dt\ket{0}_f.
\end{equation}
This is the continuous-mode version of the single-mode single-photon state 
$\ket{1}=a^*\ket{0}$ where $a^*$ is the single-mode creation operator and 
$\ket{0}$ is the ground state. 
Note that $\mbox{}_f\pro{1_\xi}{1_\xi}_f=1$ holds due to the normalization 
condition of $\xi(t)$. 
Also from the relation $\mbox{}_f\bra{1_\xi}b^*(t)b(t)\ket{1_\xi}_f=|\xi(t)|^2$, 
$\xi(t)$ has the meaning of the wave function such that $|\xi(t)|^2$ 
represents the probability of photo detection per unit time.


\subsection{Zeros of a passive linear system}

The transfer function matrix of the system \eqref{memory dynamics} is 
given by 
\[
     G(s)=[I-C(sI-A)^{-1}C^\dagger]S. 
\]
Here we give two definitions of zeros of a general $m\times m$ transfer 
function matrix $G(s)$ \cite{Zhou Doyle book}. 

{\it Definition 1:} 
If there exist $z\in{\mathbb C}$ and $\vecu\in{\mathbb C}^m$ such that 
$G(z)\vecu=0$, then $z$ is called a {\it transmission zero}. 

{\it Definition 2:} 
If there exists $z\in{\mathbb C}$ such that $G(z)=0$, then $z$ is called 
a {\it blocking zero}.

The following three facts are used in this paper. 
\\

{\it 
Fact 1: Suppose that $z\in{\mathbb C}$ is a (blocking or transmission) zero of 
$G(s)$, and it is not a pole of $G(s)$. 
Then $z$ is an eigenvalue of $-A^\dagger$. 
}

{\it Proof:} 
Let us first consider the case of a blocking zero. 
This means there exists $z\in{\mathbb C}$ such that 
$[I-C(zI-A)^{-1}C^\dagger]S=0$. 
Now let us define $V:=(zI-A)^{-1}C^\dagger S$; 
then we have $(zI-A)VS^\dagger=C^\dagger$ and $CV=S$. 
These two equations lead to $(zI-A)V=C^\dagger CV$ and therefore 
$(A+C^\dagger C)V=zV$. 
Thus $-A^\dagger V=zV$; note that $z$ is degenerated in the eigenspace 
${\rm span}(V)$. 

The case of transmission zero is almost the same. 
The definition is that there exist $z\in{\mathbb C}$ and 
$\vecu\in{\mathbb C}^m$ such that 
$G(s)\vecu=[I-C(zI-A)^{-1}C^\dagger]S\vecu=0$. 
Again define $V:=(zI-A)^{-1}C^\dagger S$, which leads to 
$(zI-A)VS^\dagger=C^\dagger$ and $CV\vecu=S\vecu$; hence we have 
$(zI-A)V\vecu=C^\dagger CV\vecu$ and $-A^\dagger V\vecu = zV\vecu$. 
Note from this the transmission-zero vector $\vecu$ and the 
eigenvector of $-A^\dagger$, $\vecv$, are connected by $\vecv=V\vecu$. 
This further yields $S^\dagger C\vecv=S^\dagger CV\vecu=S^\dagger S\vecu=\vecu$, 
hence $\vecu=S^\dagger C\vecv$. 
$\Box$
\\

{\it 
Fact 2: If $A$ is Hurwitz, then all (blocking or transmission) zeros of $G(s)$ 
are unstable zeros. 
}

{\it Proof:} 
Let $\lambda$ be an eigenvalue of $A$, i.e. ${\det}(\lambda I-A)=0$. 
This yields ${\det}(\lambda^* I-A^\dagger)=0$. 
Then from Fact 1, a zero of $G(s)$, z, is given by $z=-\lambda^*$. 
Hence ${\rm Re}(z)=-{\rm Re}(\lambda)>0$. 
$\Box$
\\

{\it 
Fact 3: If $A$ is Hurwitz, $G(s)$ always has a transmission zero. 
}

{\it Proof:} 
We begin with the eigen-equation $-A^\dagger \vecv =z\vecv$. 
Then from $-A^\dagger=A+C^\dagger C$, we have $(A+C^\dagger C)\vecv =z\vecv$. 
Now from Fact 2, $z$ is not an eigenvalue of $A$, hence $(z-A)^{-1}$ always exists; 
thus we have $\vecv=(z-A)^{-1}C^\dagger C \vecv$. 
This yields $C\vecv=C(z-A)^{-1}C^\dagger C \vecv$ and further 
$[I-C(z-A)^{-1}C^\dagger ]C\vecv=0$. 
Therefore we end up with $G(z)S^\dagger C\vecv=0$, meaning that there 
always exists a transmission zero. 
Note again we find the transmission-zero vector $\vecu$ and the eigenvector 
$\vecv$ are connected by $\vecu=S^\dagger C\vecv$. 
$\Box$


\section{General MIMO state transfer}

In what follows we assume that $A$ is Hurwitz. 
Then the solution of the dynamics is given by 
\begin{equation}
\label{general solution}
      \veca^\sharp(0) = U^*\veca^\sharp(t_0)U 
          = -\int_{t_0}^0 e^{-A^\sharp t}C^\top S^\sharp \vecb^\sharp(t)dt,
\end{equation}
where $U$ is the unitary operator describing the joint time evolution of the system 
and the field from the initial time $t_0$ to the final time $0$. 
In particular the initial time is assumed to be $t_0\rightarrow -\infty$. 
Let us define the matrix of functions 
\begin{eqnarray}
& & \hspace*{-3em}
\label{general pulse function}
    \Xi(t) = \left[ \begin{array}{ccc}
           \xi_{1,1}(t) & \cdots & \xi_{1,m}(t)  \\
           \vdots &  & \vdots  \\
           \xi_{n,1}(t) & \cdots & \xi_{n,m}(t)  \\
          \end{array} \right] 
\nonumber \\ & & \hspace*{-1.2em}
          := -e^{-A^\sharp t}C^\top S^\sharp \Theta(-t). 
\end{eqnarray}
$\Theta(-t)$ is the Heaviside step function taking $1$ for $t\leq 0$ and $0$ 
for $t>0$. 
This matrix satisfies $\int_{-\infty}^\infty \Xi(t)\Xi(t)^\dagger dt=I$. 
Then we find 
\begin{eqnarray}
& & \hspace*{-3em}
\label{entangled operators}
     U^*\veca^\sharp(t_0)U 
     = \left[ \begin{array}{c}
           U^*a_1^*(t_0)U  \\
           \vdots  \\
           U^*a_n^*(t_0)U  \\
          \end{array} \right] 
      = \int_{t_0}^0 \Xi(t) \vecb^\sharp(t)dt
\nonumber \\ & & \hspace*{2em}
      = \left[ \begin{array}{c}
           B_1^*(\xi_{1,1}) + \cdots + B_m^*(\xi_{1,m})  \\
           \vdots  \\
           B_1^*(\xi_{n,1}) + \cdots + B_m^*(\xi_{n,m})  \\
          \end{array} \right], 
\end{eqnarray}
where $B_k^*(\xi_{i,j})$ is the continuous-mode creation process operator 
on the $k$th input channel, defined by Eq.~\eqref{B and Bstar}. 
This means that a special class of  input field state can be perfectly transferred 
to the system. 
For instance let us consider the following entangled single-photon field state:
\begin{eqnarray}
& & \hspace*{-3em}
    \ket{\Psi(t_0)}_f 
        = \ket{1_{\xi_{1,1}}^{(1)}}_f + \cdots + \ket{1_{\xi_{1,m}}^{(m)}}_f
\nonumber \\ & & \hspace*{0.8em}
        = \ket{1_{\xi_{1,1}},0,\ldots,0}_f + \cdots + \ket{0,\ldots,0,1_{\xi_{1,m}}}_f 
\nonumber \\ & & \hspace*{0.8em}
        = \Big[ B_1^*(\xi_{1,1}) + \cdots + B_m^*(\xi_{1,m})\Big]
        \ket{0,\ldots,0}_f, 
\nonumber
\end{eqnarray}
where the definition \eqref{cont mode single photon} is used. 
Also we define $\ket{1^{(j)}}=\ket{0,\ldots,1,\ldots, 0}$ with $1$ appearing 
only in the $j$th component. 
Note $\mbox{}_f\pro{\Psi(t_0)}{\Psi(t_0)}_f=1$ due to 
$\int_{-\infty}^\infty \Xi(t)\Xi(t)^\dagger dt=I$. 
In this case, from Eq.~\eqref{entangled operators}, the final state of the 
whole system is calculated as 
\begin{eqnarray}
& & \hspace*{-2em}
      \ket{\Psi(0)}=U\ket{0,\ldots,0}_s\ket{\Psi(t_0)}_f 
\nonumber \\ & & \hspace*{-1.2em}
       = U\Big[ B_1^*(\xi_{1,1}) + \cdots + B_m^*(\xi_{1,m})\Big]
        \ket{0,\ldots,0}_s\ket{0,\ldots,0}_f 
\nonumber \\ & & \hspace*{-1.2em}
       = U\Big[ B_1^*(\xi_{1,1}) + \cdots + B_m^*(\xi_{1,m})\Big]U^*
        \ket{0,\ldots,0}_s\ket{0,\ldots,0}_f 
\nonumber \\ & & \hspace*{-1.2em}
       = a_1^*(t_0)\ket{0,\ldots,0}_s\ket{0,\ldots,0}_f 
       = \ket{1,0,\ldots,0}_s\ket{0,\ldots,0}_f
\nonumber \\ & & \hspace*{-1.2em}
       = \ket{1^{(1)}}_s\ket{0,\ldots,0}_f. 
\nonumber
\end{eqnarray}
This equation shows that the first mode of the system acquires the single 
photon from the field; i.e. perfect state transfer is realized.

More generally, if the input field state is given by 
\begin{eqnarray}
\label{general field state}
& & \hspace*{-3em}
    \ket{\Psi(t_0)}_f 
        = x_1\Big(\ket{1_{\xi_{1,1}}^{(1)}}_f + \cdots + \ket{1_{\xi_{1,m}}^{(m)}}_f\Big)
\nonumber \\ & & \hspace*{2.1em}
        + \cdots 
        + x_n\Big(\ket{1_{\xi_{n,1}}^{(1)}}_f + \cdots + \ket{1_{\xi_{n,m}}^{(m)}}_f\Big)
\nonumber \\ & & \hspace*{0.8em}
        = \ket{1_{\xi'_{1}}^{(1)}}_f + \cdots + \ket{1_{\xi'_{m}}^{(m)}}_f,
\end{eqnarray}
where $\xi'_{j}=x_1\xi_{1,j}+\cdots + x_n\xi_{n,j}$ and $x_j\in{\mathbb C}$ an 
arbitrary coefficient satisfying $\sum_{j=1}^n|x_j|^2=1$, in this case 
the final state is 
\begin{equation}
\label{general system state}
    \ket{\Psi(0)} 
     = \Big(x_1\ket{1^{(1)}}_s + \cdots + x_n\ket{1^{(n)}}_s \Big)\ket{0,\ldots,0}_f. 
\end{equation}
Note that the pulse functions $\xi_{i,j}(t)$ do not depend on the (unknown) 
coefficients $\{x_j\}$; 
hence, if the single photon field state with classical information $\{x_j\}$ can 
be prepared, which is a challenging task experimentally, then it can be 
perfectly transferred to the system. 
\\

{\it Example 1:} 
Let us consider the case where the system is composed of two single-mode 
SISO subsystems specified by the system parameters $(A_1, C_1)$ and 
$(A_2, C_2)$. 
(Thus $A_1, A_2, C_1, C_2$ are scalars.) 
These two subsystems can be placed at a distant location from a source. 
The two input fields are combined at a beam splitter before being sent to 
the two subsystems. 
Thus the whole system are specified by 
\[
       A= \left[ \begin{array}{cc}
           A_1 & 0  \\
           0 & A_2  \\
          \end{array} \right], ~~~
       C= \left[ \begin{array}{cc}
           C_1 & 0  \\
           0 & C_2  \\
          \end{array} \right], ~~~
       S= \left[ \begin{array}{cc}
           \alpha & \beta  \\
           \beta & -\alpha  \\
          \end{array} \right]. 
\]
Here $\alpha$ and $\beta$ represent the transmissivity and the reflectivity 
of the beam splitter, respectively, which are assumed to be real without loss of 
generality. 
Then we have 
\begin{eqnarray}
& & \hspace*{-3em}
      \Xi(t) 
         = \left[ \begin{array}{cc}
           \xi_{1,1}(t) & \xi_{1,2}(t)  \\
           \xi_{2,1}(t) & \xi_{2,2}(t)  \\
          \end{array} \right] 
\nonumber \\ & & \hspace*{-0.9em}
         = \left[ \begin{array}{cc}
           -\alpha e^{-A_1^* t}C_1 & -\beta e^{-A_1^* t}C_1  \\
           -\beta   e^{-A_2^* t}C_2 & -\alpha e^{-A_2^* t}C_2  \\
          \end{array} \right], 
\end{eqnarray}
and  
\[
     \left[ \begin{array}{c}
           U^*a_1^*(t_0)U  \\
           U^*a_2^*(t_0)U  \\
          \end{array} \right] 
      = \left[ \begin{array}{c}
           B_1^*(\xi_{1,1}) + B_2^*(\xi_{1,2})  \\
           B_1^*(\xi_{2,1}) + B_2^*(\xi_{2,2})  \\
          \end{array} \right]. 
\]
Then if the initial field state is prepared as 
\begin{eqnarray}
\label{example 1 field state}
& & \hspace*{-3em}
    \ket{\Psi(t_0)}_f 
        = x_1\Big(\ket{1_{\xi_{1,1}},0}_f + \ket{0,1_{\xi_{1,2}}}_f\Big)
\nonumber \\ & & \hspace*{4em}
       \mbox{}
        + x_2\Big(\ket{1_{\xi_{2,1}},0}_f + \ket{0,1_{\xi_{2,2}}}_f\Big)
\nonumber \\ & & \hspace*{0.8em}
        = \ket{1_{x_1\xi_{1,1}+x_2\xi_{2,1}},0}_f 
                   + \ket{0,1_{x_1\xi_{1,2}+x_2\xi_{2,2}}}_f,
\end{eqnarray}
the final system state is given by 
\[
      \ket{\Psi(0)}_s=x_1\ket{1,0}_s+x_2\ket{0,1}_s.
\]
That is, the two separately placed two subsystems are entangled. 
Note however that, to achieve the perfect state transfer, in general, the initial 
field state has to be entangled between the two input channels even before 
entering into the beam splitter. 
$\Box$

\begin{figure}
\centering
\includegraphics[scale=0.4]{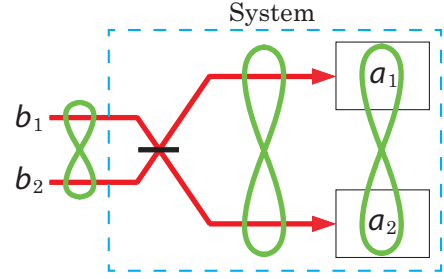}
\caption{\label{Example 1}
Perfect state transfer from the field single-photon entangled state to the 
system single-photon entangled state. 
}
\end{figure}


\section{Perfect state transfer and zeros}

In the previous section we found that, in the general setup, the field state 
\eqref{general field state} can be perfectly transferred to the system state 
\eqref{general system state}. 
That is, although engineering the entangled single-photon state 
\eqref{general field state} is challenging in experiment, perfect state 
transfer is in principle {\it always} possible. 
This can be understood in terms of systems and control theory as follows. 
In general, for a linear system if the input is of the form $\vecu(t)=\vecu e^{zt}$ 
with $z$ a transmission zero and $\vecu$ the corresponding transmission-zero 
vector, then the output is given by $\vecy(t)=G(z)\vecu e^{zt}=0$ for all $t\leq 0$. 
That is, if the system has a transmission (or more strongly blocking) zero, 
then an appropriately chosen input can make the output always zero. 
The point here is that the passive linear quantum system always has a 
transmission zero as shown in Fact~3, and this is the reason why the field 
state \eqref{general field state} is perfectly absorbed into the system. 
Thus the questions arising here are how the pulse function can be represented 
in terms of zeros of the system, and what field state represented in terms of 
zeros can be perfectly transferred.

To answer these questions let us recall Facts~1 and 3. 
That is, the system always has a transmission zero $z$ satisfying $G(z)\vecu=0$ 
with $\vecu$ the corresponding transmission-zero vector, and this satisfies the 
eigen-equation $-A^\dagger \vecv=z\vecv$ with $\vecv$ the corresponding 
eigenvector. 
Further, $\vecu$ and $\vecv$ are related as $\vecu=S^\dagger C\vecv$. 
Therefore, from Eq.~\eqref{general solution} we have
\begin{eqnarray}
\label{u-v relation}
& & \hspace*{-3em}
      \vecv^\top \veca^\sharp(0) 
         =  -\int_{t_0}^0 \vecv^\top e^{-A^\sharp t}C^\top S^\sharp \vecb^\sharp(t)dt
\nonumber \\ & & \hspace*{0.9em}
         = -\int_{t_0}^0 e^{z t} \vecv^\top C^\top S^\sharp \vecb^\sharp(t)dt
\nonumber \\ & & \hspace*{0.9em}
         = -\int_{t_0}^0 e^{z t} \vecu^\top\vecb^\sharp(t)dt
\nonumber \\ & & \hspace*{0.9em}
         = \vecu^\top \vecB^\sharp(-e^{z t})
\nonumber \\ & & \hspace*{0.9em}
         = u_1 B_1^*(-e^{z t}) + \cdots + u_m B_m^*(-e^{z t}), 
\end{eqnarray}
where we have defined 
\[
      \vecB^\sharp(\xi) = [B_1^*(\xi), \ldots, B_m^*(\xi)]^\top. 
\]
Note that $-e^{z t}$ is not normalized, but let us keep this unnormalized 
pulse function to explicitly see the transmission zero $z$. 
We here chose the following field input state:
\begin{eqnarray}
& & \hspace*{-1em}
      \ket{\Psi(t_0)}_f = \vecu^\top \vecB^\sharp(-e^{z t}) \ket{0,0,\ldots,0}_f
\nonumber \\ & & \hspace*{-0em}
        = \Big[ u_1 B_1^*(-e^{z t}) + \cdots + u_m B_m^*(-e^{z t}) \Big]
             \ket{0,0,\ldots,0}_f
\nonumber \\ & & \hspace*{-0em}
        = u_1\ket{1_{-e^{z t}}^{(1)}}_f + \cdots + u_m\ket{1_{-e^{z t}}^{(m)}}_f.
\nonumber
\end{eqnarray}
Then the final system-field state is given by 
\begin{eqnarray}
& & \hspace*{-3em}
      \ket{\Psi(0)}=U\ket{0,\ldots,0}_s\ket{\Psi(t_0)}_f 
\nonumber \\ & & \hspace*{-0em}
       = \vecu^\top U\vecB^\sharp(-e^{z t}) U^*
        \ket{0,\ldots,0}_s\ket{0,\ldots,0}_f 
\nonumber \\ & & \hspace*{-0em}
       =\vecv^\top \veca^\sharp(t_0) \ket{0,\ldots,0}_s\ket{0,\ldots,0}_f 
\nonumber \\ & & \hspace*{-0em}
       = \Big[v_1\ket{1^{(1)}}_s + \cdots + v_n\ket{1^{(n)}}_s \Big]\ket{0,\ldots,0}_f. 
\nonumber
\end{eqnarray}
Hence certainly the input pulse function needs to be of the rising exponential 
form specified by the transmission zero $z$, in order to achieve the perfect state 
transfer. 
In particular, the input field state has coefficients specified by $\vecu$, and 
the final system state has coefficients specified by $\vecv$.

The generalization is straightforward. 
Let us consider the case where the system has $\bar{m}~(\leq m)$ transmission 
zeros, $z_1, \ldots, z_{\bar{m}}$, with corresponding transmission-zero vectors 
$\vecu_1, \ldots, \vecu_{\bar{m}}$. 
Then from Eq.~\eqref{u-v relation} we have 
\begin{eqnarray}
& & \hspace*{-2.5em}
\label{entangled operators zeros}
      \left[ \begin{array}{c}
           \vecv_1^\top \veca^\sharp(0)   \\
           \vdots \\
           \vecv_{\bar{m}}^\top \veca^\sharp(0)  \\
          \end{array} \right] 
     = \left[ \begin{array}{c}
           \vecu_1^\top \vecB^\sharp(-e^{z_1 t}) \\
           \vdots \\
           \vecu_{\bar{m}}^\top \vecB^\sharp(-e^{z_{\bar{m}} t}) \\
          \end{array} \right] 
\nonumber \\ & & \hspace*{-1.5em}
     = \left[ \begin{array}{c}
           u_{1,1} B_1^*(-e^{z_1 t}) + \cdots + u_{1,m} B_m^*(-e^{z_1 t}) \\
           \vdots \\
           u_{{\bar{m}},1} B_1^*(-e^{z_{\bar{m}} t}) 
                       + \cdots + u_{{\bar{m}},m} B_m^*(-e^{z_{\bar{m}} t}) \\
          \end{array} \right], 
\end{eqnarray}
where $\vecu_j=[u_{j,1}, \ldots, u_{j,m}]^\top$. 
Now we set the field input state to be 
\begin{eqnarray}
\label{general field state zero}
& & \hspace*{-1.5em}
    \ket{\Psi(t_0)}_f 
\nonumber \\ & & \hspace*{-0.8em}
    = \Big[ x_1\vecu_1^\top \vecB^\sharp(-e^{z_1 t}) 
             + \cdots + 
               x_{\bar{m}}\vecu_{\bar{m}}^\top \vecB^\sharp(-e^{z_{\bar{m}} t}) \Big]
\nonumber \\ & & \hspace*{3.8em}
            \times
                     \ket{0,0,\ldots,0}_f
\nonumber \\ & & \hspace*{-0.8em}
        = \Big[ x_1\Big(u_{1,1} B_1^*(-e^{z_1 t}) + \cdots + u_{1,m} B_m^*(-e^{z_1 t}) \Big)
        + \cdots 
\nonumber \\ & & \hspace*{0.5em}
        \mbox{}
        + x_{\bar{m}}\Big(u_{{\bar{m}},1} B_1^*(-e^{z_{\bar{m}} t}) 
                       + \cdots + u_{{\bar{m}},m} B_m^*(-e^{z_{\bar{m}} t})\Big) \Big]
\nonumber \\ & & \hspace*{3.8em}
           \times
                     \ket{0,0,\ldots,0}_f
\nonumber \\ & & \hspace*{-0.8em}
       = \Big[ B_1^*(-x_1u_{1,1}e^{z_1 t} -\cdots 
               - x_{\bar{m}}u_{{\bar{m}},1}e^{z_{\bar{m}} t} )
                  + \cdots 
\nonumber \\ & & \hspace*{1em}
        \mbox{}
        + B_m^*(-x_1u_{1,m}e^{z_1 t} 
                -\cdots 
                     - x_{\bar{m}}u_{{\bar{m}},m}e^{z_{\bar{m}} t} ) \Big]
\nonumber \\ & & \hspace*{3.8em}
           \times
                     \ket{0,0,\ldots,0}_f
\nonumber \\ & & \hspace*{-0.8em}
       = \Big[ B_1^*(u_1')+\cdots+B_m^*(u_m') \Big]\ket{0,0,\ldots,0}_f
\nonumber \\ & & \hspace*{-0.8em}
       = \ket{1_{u_1'}^{(1)}}_f +\cdots+ \ket{1_{u_m'}^{(m)}}_f,
\end{eqnarray}
where $x_1, \ldots, x_{\bar{m}}$ are arbitrary coefficients and 
\begin{eqnarray}
& & \hspace*{-3.4em}
\label{linear comb zeros}
    \vecu'(t)
     := \left[ \begin{array}{c}
           u_1'(t)  \\
           \vdots \\
           u_m'(t)  \\
          \end{array} \right] 
\nonumber \\ & & \hspace*{-0.8em}
     = \left[ \begin{array}{c}
           -x_1u_{1,1}e^{z_1 t} -\cdots - x_{\bar{m}}u_{{\bar{m}},1}e^{z_{\bar{m}} t} \\
           \vdots \\
           -x_1u_{1,m}e^{z_1 t} -\cdots - x_{\bar{m}}u_{{\bar{m}},m}e^{z_{\bar{m}} t} \\
          \end{array} \right] 
\nonumber \\ & & \hspace*{-0.8em}
     = -x_1\vecu_1e^{z_1 t} -\cdots - x_{\bar{m}}\vecu_{\bar{m}}e^{z_{\bar{m}} t}. 
\end{eqnarray}
Then by defining the vector 
\begin{equation}
\label{general coefficient zero}
     \vecv': = x_1\vecv_1 + \cdots + x_{\bar{m}}\vecv_{\bar{m}},
\end{equation}
we find that 
\begin{eqnarray}
& & \hspace*{-3.4em}
  \vecv'\mbox{}^\top\veca^\sharp(0)
       = [x_1\vecv_1^\top + \cdots + x_{\bar{m}}\vecv_{\bar{m}}^\top]\veca^\sharp(0)
\nonumber \\ & & \hspace*{0.8em}
        = x_1\vecu_1^\top \vecB^\sharp(-e^{z_1 t}) 
             + \cdots + 
               x_{\bar{m}}\vecu_{\bar{m}}^\top \vecB^\sharp(-e^{z_{\bar{m}} t})
\nonumber
\end{eqnarray}
and thus the final system-field state is given by 
\begin{eqnarray}
& & \hspace*{-3em}
\label{general system state zero}
      \ket{\Psi(0)}=U\ket{0,\ldots,0}_s\ket{\Psi(t_0)}_f 
\nonumber \\ & & \hspace*{-0em}
       = \Big[ x_1\vecu_1^\top U\vecB^\sharp(-e^{z_1 t})U^* 
             + \cdots 
\nonumber \\ & & \hspace*{0.5em}
          \mbox{}
              + x_{\bar{m}}\vecu_{\bar{m}}^\top U\vecB^\sharp(-e^{z_{\bar{m}} t})U^* \Big]
        \ket{0,\ldots,0}_s\ket{0,\ldots,0}_f 
\nonumber \\ & & \hspace*{-0em}
       =\vecv'\mbox{}^\top \veca^\sharp(t_0) \ket{0,\ldots,0}_s\ket{0,\ldots,0}_f 
\nonumber \\ & & \hspace*{-0em}
       = \Big[v'_1\ket{1^{(1)}}_s + \cdots + v'_n\ket{1^{(n)}}_s \Big]\ket{0,\ldots,0}_f. 
\end{eqnarray}
Summarizing, if the field input state is given by Eq.~\eqref{general field state zero} 
with pulse functions \eqref{linear comb zeros}, then it is perfectly transferred 
to the system state given by Eq.~\eqref{general system state zero} with 
coefficient \eqref{general coefficient zero}; 
again, $\{\vecu_j\}$ are the transmission-zero vectors and $\{\vecv_j\}$ are the 
corresponding eigenvectors of $-A^\dagger$. 
Note that, if we {\it formally} input Eq.~\eqref{linear comb zeros} to the 
associated classical system with transfer function matrix $G(s)$, then the 
corresponding formal output is given by 
\[
     \vecy = -x_1G(z_1)\vecu_1e^{z_1 t} -\cdots 
              - x_{\bar{m}}G(z_{\bar{m}})\vecu_{\bar{m}}e^{z_{\bar{m}} t}=0.
\]
However, this does not mean that the input field state can be set for instance 
to the separable one $\ket{1_{u_1'}}_f\otimes \cdots \otimes \ket{1_{u_m'}}_f$; 
the input state we need to prepare is the entangled state 
\eqref{general field state zero}. 
\\

{\it Example 2 (continued from Example 1):} 
The transfer function matrix is given by 
\begin{eqnarray}
& & \hspace*{-3.1em}
     G(s) = \left[ \begin{array}{cc}
                  G_1(s) & 0  \\
                  0 & G_2(s)  \\
                \end{array} \right]
                \left[ \begin{array}{cc}
                  \alpha & \beta  \\
                  \beta & -\alpha  \\
                \end{array} \right]
\nonumber \\ & & \hspace*{-0.8em}
            = \left[ \begin{array}{cc}
                  \alpha G_1(s) & \beta G_1(s)  \\
                  \beta G_2(s) & -\alpha G_2(s)  \\
                \end{array} \right],
\nonumber 
\end{eqnarray}
where 
\begin{eqnarray}
& & \hspace*{-3.1em}
      G_j(s)=1-C_j(s-A_j)^{-1}C_j^\dagger=1-\frac{|C_j|^2}{s-A_j}
\nonumber \\ & & \hspace*{-0.3em}
       = \frac{s-A_j-|C_j|^2}{s-A_j}.
\nonumber 
\end{eqnarray}
Again note that $(A_j, C_j)$ are scalars. 
Clearly $G_j(s)$ has a zero $z_j=A_j+|C_j|^2$. 
Here we assume that the two subsystems are different and as a result they 
have two different zeros, i.e. $z_1\neq z_2$; but note $G_1(z_1)=0$ and 
$G_2(z_2)=0$. 
In this case, the transmission-zero vector corresponding to $z_1$ is 
given by $\vecu_1=[\alpha, \beta]^\top$, and also 
$\vecu_2=[\beta, -\alpha]^\top$ for the case $z_2$; 
\begin{eqnarray}
& & \hspace*{0em}
     G(z_1)\vecu_1 = 
        \left[ \begin{array}{cc}
                  0 & 0  \\
                  \beta G_2(z_1) & -\alpha G_2(z_1)  \\
                \end{array} \right]
         \left[ \begin{array}{c}
                  \alpha \\
                  \beta \\
                \end{array} \right]=0,
\nonumber \\ & & \hspace*{0em}
      G(z_2)\vecu_2 = 
        \left[ \begin{array}{cc}
                  \alpha G_1(z_2) & \beta G_1(z_2)  \\
                  0 & 0  \\
                \end{array} \right]
         \left[ \begin{array}{c}
                  \beta \\
                  -\alpha \\
                \end{array} \right]=0.
\nonumber 
\end{eqnarray}
The corresponding eigenvector $\vecv_1$ is given, from the proof of Fact 1, by 
\begin{eqnarray}
& & \hspace*{-1.34em}
     \vecv_1=V_1 \vecu_1 
       = (z_1-A)^{-1}C^\dagger S \vecu_1
\nonumber \\ & & \hspace*{0em}
       = \left[ \begin{array}{cc}
                  z_1-A_1 & 0  \\
                  0 & z_1-A_2 \\
                \end{array} \right]^{-1}
           \left[ \begin{array}{cc}
                  C_1^* & 0  \\
                  0 & C_2^* \\
                \end{array} \right]
\nonumber \\ & & \hspace*{2em}
     \times
           \left[ \begin{array}{cc}
                  \alpha & \beta  \\
                  \beta & -\alpha  \\
                \end{array} \right]
           \left[ \begin{array}{c}
                  \alpha \\
                  \beta \\
                \end{array} \right]
\nonumber \\ & & \hspace*{0em}
          = \left[ \begin{array}{c}
                  1/C_1 \\
                  0 \\
                \end{array} \right],
\nonumber 
\end{eqnarray}
and also $\vecv_2=[0, 1/C_2]^\top$. 
Note that these are certainly eigenvectors of $-A^\dagger$. 
Hence the input field state can be prepared to 
$\ket{\Psi(t_0)}_f=\ket{1_{u_1'},0}_f + \ket{0,1_{u_2'}}_f$ with pulse function 
\begin{eqnarray}
& & \hspace*{0em}
      \vecu'(t)
     = \left[ \begin{array}{c}
           u_1'(t)  \\
           u_2'(t)  \\
          \end{array} \right] 
       = -x_1\vecu_1 e^{z_1 t}
         -x_2\vecu_2 e^{z_2 t}
\nonumber \\ & & \hspace*{2.4em}
     = -x_1\left[ \begin{array}{c}
                  \alpha \\
                  \beta \\
                \end{array} \right]e^{z_1 t}
         -x_2\left[ \begin{array}{c}
                  \beta \\
                  -\alpha \\
                \end{array} \right]e^{z_2 t},
\nonumber 
\end{eqnarray}
and the system final state is then given by 
$\ket{\Psi(0)}_s = v'_1\ket{1,0}_s + v'_2\ket{0,1}_s$ with 
$\vecv'=x_1\vecv_1+x_2\vecv_2$. 
$\Box$


\section{Separable input field}

As mentioned before, the input field state (e.g. Eq.~\eqref{general field state}) 
is in general entangled among input channels and is not always easy to 
generate in experiment. 
Hence it is reasonable to seek some conditions for perfect state transfer such that 
the input field state can be prepared relatively easily; 
in particular here we focus on a {\it separable state} such as 
$\ket{1_{e^{zt}},0,\ldots,0}_f$.

The first condition is, as expected, that the system has a blocking zero $z$. 
In this case, as seen in Eq.~\eqref{u-v relation}, $\veca^*(0)$ can be represented 
in terms of $z$ as follows; that is, using the relation $CV=S$ found in the 
proof of Fact~1, Eq.~\eqref{general solution} yields 
\begin{eqnarray}
\label{separated operators zeros}
& & \hspace*{-2.5em}
      V^\top \veca^\sharp(0) 
         =  -\int_{t_0}^0 V^\top e^{-A^\sharp t}C^\top S^\sharp \vecb^\sharp(t)dt
\nonumber \\ & & \hspace*{1.6em}
         = -\int_{t_0}^0 e^{z t} V^\top C^\top S^\sharp \vecb^\sharp(t)dt
         = -\int_{t_0}^0 e^{z t} \vecb^\sharp(t)dt
\nonumber \\ & & \hspace*{1.6em}
        = \vecB^\sharp(-e^{z t})
        = \left[ \begin{array}{c}
           B_1^*(-e^{z t})  \\
           \vdots \\
           B_m^*(-e^{z t})  \\
          \end{array} \right]. 
\end{eqnarray}
Hence by introducing the normalized pulse function $\zeta(t)=-\sqrt{z+z^*}e^{zt}$, 
which satisfies $\int_{-\infty}^0|\zeta(t)|^2dt=1$, we have 
\[
        \sqrt{z+z^*} V^\top \veca^\sharp(0) = \vecB^\sharp(\zeta). 
\]
A remarkable feature of this relation is that the field operators are 
``disentangled", unlike Eqs.~\eqref{entangled operators} and 
\eqref{entangled operators zeros} which has the form of entangled operators, 
$X_1\otimes I \otimes \cdots \otimes I + \cdots + 
I\otimes \cdots \otimes I \otimes X_m$. 
This means that a separable input field state can be chosen. 
For instance, let us consider 
\[
      \ket{\Psi(t_0)}_f = B_1^*(\zeta)\ket{0,0,\ldots,0}_f
         = \ket{1_{\zeta},0,\ldots,0}_f.
\]
Then the system final state is given by 
\begin{eqnarray}
& & \hspace*{-2.4em}
      \ket{\Psi(0)}_s = \vecv_1^\top \veca^\sharp(t_0) \ket{0,0,\ldots,0}_s
\nonumber \\ & & \hspace*{1em}
        = v_{1,1}\ket{1,0,\ldots,0}_s + \cdots + v_{1,n}\ket{0,0,\ldots,1}_s, 
\nonumber
\end{eqnarray}
where $\vecv_1=[v_{1,1},\ldots,v_{1,n}]^\top$ is the first column vector of 
$\sqrt{z+z^*}V$. 
That is, if the system has a blocking zero, then a separable field state can be 
used to achieve the perfect state transfer. 
\\

\begin{figure}
\centering
\includegraphics[scale=0.4]{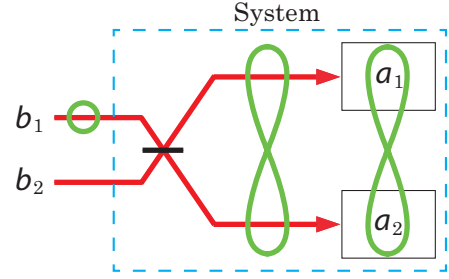}
\caption{\label{Example 3}
Schematic for entanglement creation and distribution, which stems from 
a single photon field state with engineered pulse shape. 
}
\end{figure}

{\it Example 3 (continued from Example 2):} 
If the two subsystems are identical, i.e. $A_1=A_2$, $C_1=C_2$, then the 
two transfer functions become equal, $G_1(s)=G_2(s)$. 
Hence the whole transfer function matrix is given by 
\[
     G(s) = \left[ \begin{array}{cc}
                  \alpha G_1(s) & \beta G_1(s)  \\
                  \beta G_1(s) & -\alpha G_1(s)  \\
                \end{array} \right].
\]
Clearly in this case the system has a blocking zero, $z$, satisfying $G_1(z)=0$, 
which is equal to $z=A_1+|C_1|^2$. 
Then the $V$ matrix appearing in Eq.~\eqref{separated operators zeros} 
is given by 
\begin{eqnarray}
& & \hspace*{-1.4em}
      V = (z-A)^{-1}C^\dagger S 
\nonumber \\ & & \hspace*{-0.3em}
             = \left[ \begin{array}{cc}
                  z_1-A_1 & 0  \\
                  0 & z_1-A_1 \\
                \end{array} \right]^{-1}
           \left[ \begin{array}{cc}
                  C_1^* & 0  \\
                  0 & C_1^* \\
                \end{array} \right]
           \left[ \begin{array}{cc}
                  \alpha & \beta  \\
                  \beta & -\alpha  \\
                \end{array} \right]
\nonumber \\ & & \hspace*{-0.3em}
          = \frac{1}{C_1}
              \left[ \begin{array}{cc}
                  \alpha & \beta  \\
                  \beta & -\alpha  \\
                \end{array} \right].
\nonumber
\end{eqnarray}
Therefore, if we prepare the field initial state as 
\[
      \ket{\Psi(t_0)}_f = B_1^*(\zeta)\ket{0,0}_f
         = \ket{1_{\zeta},0}_f,
\]
then the system final state is given by 
\[
      \ket{\Psi(0)}_s = \alpha\ket{1,0}_s + \beta\ket{0,1}_s, 
\]
where we have used the fact that $|\sqrt{z+z^*}/C_1|=1$.


Note again that in this case we only need to prepare a single photon field 
state living in one channel; 
then this state becomes entangled after being combined at the beam splitter, 
and further it is perfectly transferred to the two identical systems, which 
can be spatially separated as shown in Fig.~\ref{Example 3}. 
That is, the schematic proposed here can be used for the purpose of creating 
and distributing entanglement in a quantum network. 
Specifically, it can be applied for constructing quantum repeaters 
\cite{Repeater review} to realize a long-distance quantum communication. 
$\Box$
\\

Another condition such that a separable field input state can be perfectly 
transferred is as follows; if the transmission-zero vector $\vecu$ appearing 
in Eq.~\eqref{u-v relation} is e.g. of the form 
$\vecu=[1, 0, \ldots, 0]^\top$, then Eq.~\eqref{u-v relation} gives 
\[
        \vecv^\top\veca^\sharp(0)=B_1^*(-e^{zt}),
\]
or equivalently $-\sqrt{z+z^*}\vecv^\top\veca^\sharp(0)=B_1^*(\zeta)$ with 
the normalized rising exponential function $\zeta(t)=-\sqrt{z+z^*}e^{zt}$. 
In this case, the separable initial field state 
$\ket{\Psi(t_0)}_f=\ket{1_\zeta,0,\ldots,0}_f$ 
can be perfectly transferred to the system and the final system state is 
$\ket{\Psi(0)}_s=v'_1\ket{1^{(1)}}_s + \cdots + v'_1\ket{1^{(n)}}_s$ with 
$v_j'$ the $j$th component of the vector $-\sqrt{z+z^*}\vecv$. 
\\

\begin{figure}
\centering
\includegraphics[scale=0.42]{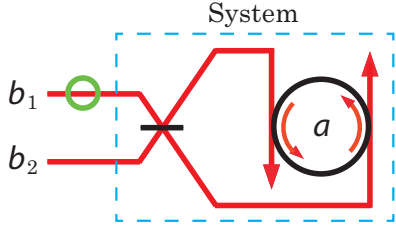}
\caption{\label{Example 4}
Single photon absorption into the single-mode ring-resonator with 
two wave guides. 
}
\end{figure}

{\it Example 4:} 
Let us consider the system studied in \cite{Huang Agarwal 2014}, depicted in 
Fig.~\ref{Example 4}. 
The system is a single-mode ring resonator coupled to two optical waveguides, 
hence it is a 2-input and 2-output system. 
The wave guides are combined at a beam splitter before connected to the 
resonator. 
The transfer function of this system is given by 
\begin{eqnarray}
& & \hspace*{-1.4em}
      G(s) = 
      \frac{1}{s+(\gamma_1+\gamma_2)/2}
\nonumber \\ & & \hspace*{-0.3em}
   \times
       \left[ \begin{array}{cc}
           s + (\gamma_2-\gamma_1)/2   &   -\sqrt{\gamma_1\gamma_2}   \\
           -\sqrt{\gamma_1\gamma_2}   &   s + (\gamma_1-\gamma_2)/2   \\
                \end{array} \right]
       \left[ \begin{array}{cc}
           \alpha   &   \beta     \\
           \beta     &   -\alpha  \\
                \end{array} \right], 
\nonumber
\end{eqnarray}
where $\gamma_1$ and $\gamma_2$ are coupling constants between the 
resonator and the waveguides. 
Also $\alpha$ and $\beta$ are the transmissivity and the reflectivity of the 
beam splitter, respectively. 
Clearly $G(s)$ does not have a blocking zero, but (as guaranteed by Fact~3) 
it has a transmission zero $z=(\gamma_1+\gamma_2)/2$ with corresponding 
transmission-zero vector 
\[
      \vecu 
            = \left[ \begin{array}{cc}
                 \alpha   &   \beta     \\
                 \beta     &   -\alpha  \\
               \end{array} \right]
               \left[ \begin{array}{c}
                        \sqrt{\gamma_1}  \\
                        \sqrt{\gamma_2}  \\
                \end{array} \right]
            = \left[ \begin{array}{c}
                        \alpha\sqrt{\gamma_1}+\beta\sqrt{\gamma_2}  \\
                        \beta\sqrt{\gamma_1}-\alpha\sqrt{\gamma_2}  \\
                \end{array} \right].
\]
Therefore from the result of Section IV we need to prepare the following 
(unnormalized) entangled input field state: 
\begin{eqnarray}
& & \hspace*{-1.4em}
       \ket{\Psi(t_0)}_f = 
           (\alpha\sqrt{\gamma_1}+\beta\sqrt{\gamma_2})\ket{1_{e^{zt}},0}_f 
\nonumber \\ & & \hspace*{3em}
        \mbox{}
             + (\beta\sqrt{\gamma_1}-\alpha\sqrt{\gamma_2})\ket{0,1_{e^{zt}}}_f, 
\nonumber
\end{eqnarray}
to achieve the perfect state transfer. 
However, in the special case where the parameters satisfy the condition 
$\beta\sqrt{\gamma_1}-\alpha\sqrt{\gamma_2}=0$, which leads to 
$\vecu=[1,0]^\top$, we only need to prepare a separable input field state 
$\ket{\Psi(t_0)}_f = \ket{1_{e^{zt}},0}_f$, and it is perfectly transferred to 
the system (see Fig.~\ref{Example 4}). 
Note that in this case, because the system is single-mode, the final system 
state is merely $\ket{1}_s$. 
$\Box$



\section{Conclusion}

In this paper, we first showed that the MIMO passive linear system always has 
a transmission zero, which ensures that a field single-photon state with 
appropriately engineered pulse function can be perfectly transferred to the 
system. 
Although in general the field state has to be an entangled state, under 
additional specific condition, this requirement can be relaxed; that is, as proven 
in Section V, a separable field state can be perfectly transferred to the system. 
This leads to a convenient schematic for creating and distributing entanglement 
in a quantum network.




\end{document}